\documentstyle[psfig,epsfig]{acmconf}

\renewcommand{\footnotesep}{5mm}

\newcommand{\comment}[1]{}

\onecolumn
\begin{document}
\title{Comparative Analysis of Five XML Query Languages\thanks{This 
research was done when the authors
were visiting Stanford University.
It was supported by ESPRIT Project 28771 W3I3, MURST Project
Interdata, CNR-CESTIA, and the HP Internet Philanthropic Initiative.}}

\author{Angela Bonifati, Stefano Ceri  \\  
{\small Dipartimento di Elettronica e Informazione, 
Politecnico di Milano} \\
{\small Piazza Leonardo da Vinci 32, I-20133 Milano, Italy} \\
{\small {\tt bonifati/ceri@elet.polimi.it}}
}          
\date{}
\maketitle

\begin{abstract}
XML is becoming the most relevant new standard for data representation
and exchange on the WWW. Novel languages for extracting and
restructuring the XML content have been proposed, some in the
tradition of database query languages (i.e. SQL, OQL),
others more closely inspired by XML.  No standard for
XML query language has yet been decided, but the discussion is ongoing
within the World Wide Web Consortium and within many academic
institutions and Internet-related major companies. We present a
comparison of five, representative query languages for XML,
highlighting their common features and differences.
\end{abstract}

\section{Introduction to the five languages}

\subsection{LOREL}

LOREL was originally designed for querying semistructured data and has
now been extended to XML data; it was conceived and implemented at
Stanford University (S. Abiteboul, D. Quass, J. McHugh, J. Widom,
J. Wiener) and its prototype is at http://www-db.stanford.edu/lore.
It is a user-friendly language in the SQL$\backslash$OQL style, it
includes a strong mechanism for type coercion and permits very
powerful path expressions, extremely useful when the structure of a
document is not known in advance \cite{AQ*97,AG*97,GMW99}.

\subsection{XML-QL}

XML-QL was designed at AT\&T Labs (A. Deutsch, M. Fernandez,
D. Florescu, A. Levy, D. Suciu);
its prototype is reachable at the url:
http://www.research.att.com/sw/tools/xmlql
as part of the Strudel Project.  The XML-QL language
extends SQL with an explicit CONSTRUCT clause for building the
document resulting from the query and uses the {\em element patterns}
(patterns built on top of XML syntax) to match data in an XML
document. XML-QL can express queries as well as transformations,
for integrating
XML data from different sources \cite{DF*98,DF*99}.  

\subsection{XML-GL}

XML-GL is a graphical query language, relying on a graphical
representation of XML documents and DTDs by means of labelled {\em XML
graphs}. It was designed at Politecnico di Milano (S. Ceri, S. Comai,
E. Damiani, P. Fraternali, S. Paraboschi and L. Tanca); an
implementation is ongoing.  All the elements of XML-GL are displayed
visually; therefore, XML-GL is suitable for supporting a user-friendly
interface (similar to QBE) \cite{C*99}.

\subsection{XSL}

The Extensible Stylesheet Language (XSL) has facilities that could serve  
as a basis for an XML query language. An XSL program consists of a 
collection of template rules; each template rule has two parts: a pattern 
which is matched against nodes in the source tree and a template which 
is instantiated to form part of the result tree. XSL makes use of the 
expression language defined by XPath \cite{Cl99b} for selecting elements 
for processing, for conditional processing and for generating text. 
It was designed by the W3C XSL working group (J. Clark editor) 
\cite{Cl99a, SLR98, W*98} . 

\subsection{XQL}

XQL is a notation for selecting and filtering the elements and text of
XML documents. XQL can be considered a natural extension to the XSL
pattern syntax; it is designed with the goal of being syntactically
very simple and compact (a query could be part of a URL), with a
reduced expressive power. It was designed by J. Robie,
Texcel Inc., J. Lapp, webMethods, Inc., and D. Schach, Microsoft
Corporation \cite{RLS98,Ro99a,Ro99b,Ro99c,SLR98}.

\subsection{Other languages} 
Several other languages for XML have been proposed, including 
XMAS \cite{LP*99} and XQuery \cite{DeR98}.

\subsection{Outline}
The paper is organized as follows: in Section~\ref{features},
we describe how the languages support various orthogonal features; in 
Section~\ref{examples},
we give some comparative examples of queries in the five languages; in  
light of these discussions,
Section~\ref{qualities} summarizes the desired qualities of the 
languages, and the concluding Section~\ref{conclusions}  
proposes a language taxonomy which shows, in a synthetic and effective way,
a comparison of the expressive power of the five languages.

\section{Features Classification}
\label{features}
In the following, we examine how the considered languages support
the various features of database query languages.
\subsubsection*{A. Data Model}
\begin{enumerate}
\item {\em Specific data model.}  The designers of LOREL, XML-QL,
XML-GL and XSL have felt the need of introducing their own data model
explicitly, while XQL relies on the data modeling features of XML.
\begin{verbatim}
LOREL: Yes    XML-QL: Yes    XML-GL: Yes    XSL:Yes       XQL: No
\end{verbatim}
In the LOREL data model, an XML element is a pair $<eid, value>$,
where $eid$ is a unique {\em element identifier}, and {\em value} is
either an atomic text string or a complex value containing a
string-valued {\em tag}, followed by a (possibly empty) ordered list
of pairs of attribute names and atomic values (representing XML
attributes), followed by a (possibly empty) ordered list of pairs
$<eid, value>$ called
{\em crosslink subelements} (representing XML IDREF
attributes), followed by a (possibly empty) ordered list of 
pairs $<eid, value>$ called {\em normal subelements} (representing the XML
containment relationship).  A graph representation of the data model
is provided; nodes correspond to the XML data elements and edges are
designed as either {\em crosslink edges} or {\em normal edges}; edges are
labeled. Each XML graph has one or more node designed as {\em entry points}.

In XML-QL, an XML document is modeled by an {\em XML Graph}; each node
is associated to an {\em object identifier} (OID); edges are labeled
with element tag identifiers, intermediate nodes are labeled with sets
of attribu\-te-value pairs representing attributes, leaves are labeled
with values (e.g., CDATA or PCDATA); each graph has a distinguished
node called the {\em root}.

In XML-GL, an {\em XML Graphical Data Model (XML-GDM)} is used to
represent both XML DTDs and actual documents; XML-GDM is also used for
formulating queries. XML elements are represented as rectangles and
properties as circles; these include attributes (as pairs of label and
value) as well as printable values CDATA or PCDATA (with value but no
label); IDREFs are denoted by black circles.  Edges between nodes
represent containment or reference relationships.

XSL operates on source, result and stylesheet documents using the same 
data model, defined in the XPath specification. An XML document is 
modeled as a tree that contains seven types of nodes (root nodes, 
element nodes, text nodes, attribute nodes, 
namespace nodes, processing instruction nodes and comment nodes). For 
every type of node, there is a way of determining a {\em string value} 
for it, that is either part of the node or computed 
from the string-values of its descendants. Some types of 
node also have an expanded name, which is a pair consisting of a local 
part and a namespace URI.

XQL designers assume the ``XML implied data model'', highlighting that
each node has a type and either content or a value, and that semantic
relationships between nodes can be hierachical (parent/child, 
ancestor/descendant),
positional (absolute, relative, range) and sequential (precedes,
immediately precedes) \cite{Ro99c}.

All of the data models introduced in LOREL, XML-QL, XML-GL and XSL may
consider the elements in a given XML document as either unordered or
ordered; however,
only XML-QL is able to query the ordering of the underlying data model,
as discussed in point G.3. 

The data models of LOREL, XML-QL, and XML-GL are substantially
equivalent; the only significant difference concerns the management of
IDREFS, which is discussed next. XSL data model is a tree rather than a 
graph, so it cannot be considered equivalent to the previous ones.

\item {\em Differential management of IDREFs.}  To support element
sharing, XML reserves an attribute of type ID, which associates an
unique key to the element.  An attribute of type IDREF allows an
element to refer to another element with the designed key. Thus,
IDREFs are particular strings that can be interpreted as references
between elements. With such an interpretation, it is possible to
navigate from one element to another; the data model supports {\em
object references}, possibly cyclic. However, if IDREFs are interpreted as
strings,
then nodes are connected only by containment relationships, and the
data model does not support object references. These two
interpretations of IDREFs may lead to two different interpretations of the
same query. We indicate this property as differential management of
IDREFs.
\begin{verbatim} 
LOREL: Yes    XML-QL: No    XML-GL: No    XSL: No       XQL: No
\end{verbatim}
In LOREL, there are two modes of viewing the data model: {\em
semantic} and {\em literal}. In the semantic mode, the database is
viewed as an interconnected graph; in the literal mode, the database
is viewed as an XML tree, and IDREFs are represented as textual
strings. \footnote {{\tt FollowLinks} is a command line directive which 
enables the semantic interpretation of the data model 
in the Lore implementation.}
\end{enumerate}

\subsubsection*{B. Basic Query Abstractions}
\begin{enumerate}
\item {\em Document selection.} Given a document and a query on the
document, the document selection is the result of the application of
the query to the document, picking up the elements, specified in the
select expression of the query, that satisfy the query condition.
\begin{verbatim}
LOREL: Yes    XML-QL: Yes    XML-GL: Yes    XSL: Yes      XQL: Yes
\end{verbatim}
Of course this feature is supported by all the languages. We next
introduce simplified descriptions of the syntax of the five languages.

A query in LOREL is structured according to the following, simplified
grammar (for full details, the reader can refer the \cite{AQ*97} paper)
\footnote{We use a BNF notation where nonterminals are enclosed within
quotes, curly brackets denote a list of zero or more elements
separated by commas, square brackets denote optionality, and the
symbol $|$ denotes alternatives.}:
\begin{verbatim}
`select' { select_expr }
[ `from' { from_expr } ]
[ `where' { where_expr } ]
\end{verbatim}
Select expressions, from expressions, and where expressions, 
as in OQL, can in turn contain queries.

A simplified syntax for defining a query in XML-QL is:
\begin{verbatim}
Query ::= `where' { Predicate }
          `construct' { `{` Query '}' }
\end{verbatim}
The result specified in the {\tt construct} clause is a piece of XML
document, fully specified in terms of tag names and content; content
is typically constructed from the object bindings which are
determined by the predicate evaluation (for full details on the XML-QL 
syntax, see \cite{DF*98}).

An XML-GL query consists of two sets of directed acyclic graphs
displayed side by side and separated by a vertical line, where the LHS
express the query sources (which documents are selected) and predicate
(which condition must be satisfied), while the RHS represents the
construction (which document is produced as result); explicit
connections or implicit homo\-nims, where unambiguous, indicate the
bindings between the LHS and RHS.

In XSL, a template rule is specified within the {\tt xsl:tem\-plate} tags. 
The {\tt match} attribute is a pattern that identifies the source node or 
nodes to which the rule applies. If it is omitted, the template rule is 
matched against all the nodes of the document. The following is the 
approximate skeleton of a template rule (see \cite{Cl99a} for complete 
information): 
\begin{verbatim}
`<xsl:template'  [`match=' pattern_expr] `>'
    {`<xsl:directive>'}
    {`<result-elements>'}   
`</xsl:template>' 
\end{verbatim}
where {\tt pattern$\_$expr} represents an expression written in the XPath 
language (see section C.1 for more details),       
{\tt result-el\-ements} are the new elements' tags produced in the 
result
and {\tt xsl:directive} is one of the following
(for a complete list, refer to \cite{Cl99a}):
\begin{verbatim}
`<xsl:apply-templates'[`select=' pattern_expr]`>'
`<xsl:for-each' `select=' pattern_expr `>'       
`<xsl:value-of' [`select=' pattern_expr] `>'
`<xsl:copy-of'   `select=' pattern_expr `>'
\end{verbatim}     
where the {\tt select} attribute
is used to process nodes selected by an expression instead of all the
children of the current node.
The {\tt xsl:apply-templates} directive invokes the application of 
templates which are separately defined;
the {\tt xsl:for-each} directive iterates on other directives which are 
statically nested within it;
the {\tt xsl:value-of} directive extracts information from a pattern 
expression and converts it into a string; 
the {\tt xsl:copy-of} directive extracts information from a pattern 
expression and puts it into the result, with its original format.

The basic XQL syntax mimics the URI directory navigation syntax, but
instead of specifying navigation through a physical structure, the
navigation is through elements in the XML tree. A simplified
syntax for XQL is (see \cite{RLS98} for the complete syntax):
\begin{verbatim}
Query ::= [ `./' | `/' | `//' | `.//'] 
          Element [ `[' Predicate `]' ][Path]
Path  ::= [ `/' | `//' ] Element 
          [ `[' Predicate `]' ] [Path]
\end{verbatim}
Therefore, a query is specified along a specific hierarchical path
within the document; predicates (filters in XQL terminology) typically
apply to the elements being accessed during the navigation along the
path, although syntactically they may reach elements which are quite
distant in the hierarchical structure.

\item {\em Joins.} A join condition compares two or more XML
attributes or data belonging to the same document or to two different
documents. The typical comparison (equality) is called equi-join.

\begin{verbatim}
LOREL: Yes    XML-QL: Yes    XML-GL: Yes    XSL: No       XQL: No
\end{verbatim}

In LOREL, join conditions are fully supported, within the same
document and among several documents. They are written in a SQL-like
form, by explicitly specifying the variables involved in the joins.

In XML-QL, joins are implicitly expressed by means of the equality on
variable names, which must match their value. They can range on the
same document or on several different documents, defining arbitrary
join conditions (e.g., $n$-way joins correspond to associating the
same variable name to $n$ labels in the query).

In XML-GL, join conditions are expressed by connecting, by means of a
comparison operator, two node leaves of XML documents, representing
arbitrary attributes or data; equi-joins are represented by edges
pointing to the same node, possibly with two labels (in many cases
joined nodes have the same label).

XSL does not allow neither joins nor semi-joins: each template rule 
currently addresses a single document and no connection  
conditions can be built within the same document or among several 
documents. 

XQL allows semi-joins, i.e., joins of data which is reachable along a
path with other data which may be present in the same document; it
does not allow joins among different documents. For instance, it is
possible to ask for all books whose author is the same as the author
of Moby Dick within a given document:
\begin{verbatim}
book[author=//book[title=`Moby Dick']/author]
\end{verbatim} 

\item {\em Semantics of the query result.}  The result of a query can
be defined in terms of the current content of the database (i.e., by
pointing to object identifiers of nodes in the document base) or as a
new document, which can be then queried and possibly updated
independently. In general, we expect that all languages can adapt to
either solutions. We anyway indicate the default
solution adopted by the five languages.
\begin{verbatim}
LOREL: SetOfOIDs   XML-QL: XMLDoc   XML-GL: XMLDoc  XSL: XMLDoc  XQL: XMLDoc 
\end{verbatim}

In LOREL, the result of a query is a set of object identifiers pointed
by a new element.
Therefore, the standard interpretation is that the current state of 
objects being
selected is the one present in the database, and subsequent accesses
to the query result may give different documents. In LOREL it is also
possible to define views ({\tt with} clause), and in such case the
query returns a document with all the nodes which are specified by the
{\tt with} clause.  In the standard interpretations, XML-QL, XML-GL, XSL
and XQL return new documents, whose content is then independent on
subsequent database manipulations.

\end{enumerate}

\subsubsection*{C. Path expressions}

\begin{enumerate}
\item {\em Partially specified path expressions.} When querying
semistructured data, especially when the exact structure is not known,
it is convenient to use a form of ``navigational'' query based on path
expressions. The most powerful form of path expression does not need
to list all the elements of the path, as it uses wildcards and regular
expressions: we denote it as a partially specified path expression.
All languages support partially specified path expression and actually
consider this feature as one of the most important in the language.

\begin{verbatim}
LOREL: Yes    XML-QL: Yes   XML-GL: Partially   XSL: Yes      XQL: Yes
\end{verbatim}

In LOREL, path expressions are very powerful and flexible; they admit
several {\em Unix-like wildcards}. Each path expression must have a
context (the root element of the document).

In XML-QL, path expressions are admitted within the tag specification
and they permit the alternation, concatenation and Kleene-star
operators, similar to those used in regular expressions.  XML-QL path
expressions have the same expressive power as those of LOREL.

In XML-GL, the only path expressions supported are arbitrary
containment, by means of a wildcard * as edge label; this allows
traversing the XML-GL graph reaching an element at any level of depth.

In both XSL (XPath language) and XQL, path expressions define 
relative and absolute locations. 
A relative location path
consists of one or more location steps (XML nodes) separated by the child 
`/' operator or by the descendant `//' operator. An absolute
location path has a `/' or  `//' optionally followed by a relative location 
path.
Admitted wildcards are both alternation and Kleene-star closure 
operator. 

\item{\em Matching of partially specified expressions with cyclic
data.}  Partially specified expressions may be the source of infinite
computations in the case of cyclic instances. Therefore, it is common
practice to specify halt conditions in the matching algorithm that
binds object instances to path expressions when the same object
binding is associated to the same query node more than once. 
\begin{verbatim}
LOREL: Yes  XML-QL: Undefined  XML-GL: Yes    XSL: No     XQL: Undefined
\end{verbatim} 
Some systems do not mention the halt condition as part of the query
language semantics (we expect it to be part of the implementation).

In XSL, {\tt xsl:apply-templates} are used 
to process only nodes that are descendants of the current node, and this 
cannot result in non-terminating processing 
loops. However, non-terminating loops may arise when {\tt 
xsl:ap\-ply-templates} is used to process 
elements that are not descendants of the current node. For example, the 
template rule: 
\begin{verbatim}
<xsl:template   match="foo">
  <xsl:apply-templates    select="."/>
</xsl:template>
\end{verbatim} 

matches the \verb+<FOO>+ elements at all levels of nesting, including the 
level on which the matching occurs, yielding to a possible infinite call 
sequence.  
Implementations may be able to detect such loops in some cases, but 
a stylesheet may enter a non-terminating loop, depending on the matching 
algorithm.   
\end{enumerate}

\subsubsection*{D. Quantification, Negation, and Reduction}

\begin{enumerate}
\item {\em Existential quantification.} An existential predicate over
a set of instances (e.g., bound to a variable) is satisfied if
at least one of the instances satisfies the predicate.

\begin{verbatim}
LOREL: Yes    XML-QL: Yes    XML-GL: Yes   XSL: Yes   XQL: Yes
\end{verbatim}

In all languages, predicates are assumed as existentially quantified.

\item {\em Universal quantification.} An universal predicate over a
set of instances (e.g., bound to a variable) is satisfied if all the
instances satisfies the predicate.

\begin{verbatim}
LOREL: Yes    XML-QL: No    XML-GL: No    XSL: No      XQL: Yes
\end{verbatim}

In LOREL, a variable can be universally quantified with the SQL-like
predicate {\tt for all}. Similarly, in XQL universal quantification
is obtained by prefixing a predicate expression with the keyword $\$
all\$$. In XML-QL and XML-GL\footnote{However, in XML-GL the presence of 
negation and existential quantification allows the expression of universal 
quantification by means of views.} universal quantification cannot be 
expressed.
In the XSL pattern language, conditions cannot be universally quantified.

\item {\em Negation.} The negation of a predicate over a set of
instances is satisfied if none of the instances satisfies the
predicate.

\begin{verbatim}
LOREL: Yes    XML-QL: No    XML-GL: Yes    XSL: Yes      XQL: Yes
\end{verbatim}

In LOREL, a predicate is negated with the key-word {\tt not}.
In XML-GL, the negation is expressed graphically by a dashed edge (see
Figure~\ref{exnot}, where all professors without a name element are
selected). In XQL, the boolean operator $\$not\$$ negates
the value of an expression within a filter, e.g.:
\begin{verbatim}
professor[$not$ name]
\end{verbatim}
Its meaning is to select all the instances of the context associated
with the filter with a negation predicate where the expression is
false or not valuable (e.g., the name attribute is missing). XML-QL
does not support negation (it supports the unequal comparison operator
in simple predicate expressions). In the XSL pattern language, negation is 
expressed by means of the boolean function {\tt not} that operates 
on a boolean value and returns a boolean (before using the {\tt not} 
function, a number, a string, a node set 
and an object must be first converted to their boolean value by applying the 
{\tt boolean} function to them).

\begin{figure}[htbp]
\centerline{\psfig{figure=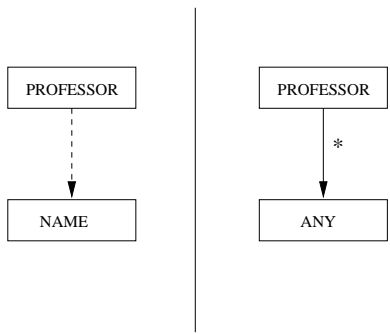,width=2in}}
\caption{Example of negation in XML-GL}
\label{exnot}
\end{figure}

\item {\em Reduction.} Given a document and a query on this document, 
the reduction
prunes from the document those elements specified in the selection
part of the query that satisfy its condition.

\begin{verbatim}
LOREL: No    XML-QL: No    XML-GL: No    XSL: No      XQL: No
\end{verbatim}

In LOREL, XML-QL, XML-GL and XSL document reduction is not supported, also if
it could be obtained through a construction which includes only the
elements that should remain.  This is feasible only when the
underlying DTD of the document is known in advance. In XQL no
construction can be expressed, and therefore this approach is not
feasible.
\end{enumerate}

\subsubsection*{E. Restructuring abstractions}
\begin{enumerate}
\item {\em Building new elements}. A new XML element can be created
through the query's construction mechanism.
\begin{verbatim}
LOREL: Yes    XML-QL: Yes    XML-GL: Yes    XSL: Yes      XQL: No
\end{verbatim}

In LOREL, a new XML element is built by invoking the {\tt
xml()} function with three parameters (the first two not mandatory): the 
type, the label 
and the value(s), explicitly given through the OID or 
implicitly given specifying the query that generate it. 

In XML-QL, restructuring is specified in the {\tt Construct} clause,
which contains the new tags, constants, and variables (bound from the
XML-QL predicate evaluation) of the new document, arbitrarily named.

In XML-GL, a new element is constructed through a new element box
arbitrarily named. New elements can be added and arbitrarily named. 

In XSL, the new elements of the query result are specified through their tag 
names within the template rule. 

In XQL, no new element can be added to the existing ones, because no
construction mechanism is provided.  

\item {\em Grouping}. Elements of the result can be aggregated or
reorganized as specified by means of special functions, such as {\tt
group by}.
\begin{verbatim}
LOREL: Yes    XML-QL: No    XML-GL: Yes    XSL: No       XQL: No
\end{verbatim}

In LOREL, the {\tt group by} clause is that inherited from the OQL
language.
Objects extracted in XML-GL can be grouped according to the distinct
values of one of their attributes or PCDATA; each class is associated
to an element instance carrying the representative value of the class
and then the rest of the XML tree as sub-element. Apparently, a
group-by clause is missing from the current descriptions of
XML-QL, XSL and XQL.

\item {\em OID invention with a Skolem function.}  A Skolem function
associates to a given value a unique, generated OID; the Skolem
function applied to the same value produces the same OID.  Skolem
functions are needed for integrating existing documents into one.
\begin{verbatim}
LOREL: Yes    XML-QL: Yes    XML-GL: Partially    XSL: No       XQL: No
\end{verbatim}
In LOREL, XML-QL and XML-GL the Skolem function takes as input a list of 
variables as
argument and returns one unique element for every binding of elements
and$\backslash$or attributes to the argument. In Lorel and XML-QL,
the Skolem function associates to each element a unique object
identifier, that can be referenced by means of variables; 
XML-GL, instead, does not 
provide the possibility of explicitly referencing the object identifiers. In 
XSL and XQL, the Skolem functions are not used. 
\end{enumerate}

\subsubsection*{F. Aggregation, Nesting, and Set operations}
\begin{enumerate}
\item {\em Aggregates.} Aggregate functions compute a scalar value out
of a multi-set of values. Classical aggregates, supported by SQL, are
{\tt min}, {\tt max}, {\tt sum}, {\tt count}, {\tt avg}.
\begin{verbatim}
LOREL: Yes   XML-QL: No  XML-GL: Yes  XSL: Partially  XQL: Partially 
\end{verbatim}

In LOREL, the aggregate functions are present and fully implemented.
In XML-QL, as of the current date they are not supported, but they are
indicated as to be supported in the next version.  In XML-GL,
aggregates are represented graphically.  In XSL and XQL, we found the method
{\tt count()} that evaluates to the number of reference nodes that
appear in the associated set. We are not aware of further aggregate
functions.

\item {\em Nesting of queries.} As in SQL, a query can be composed of
nested subqueries.
\begin{verbatim}
LOREL: Yes    XML-QL: Yes    XML-GL: No    XSL: Yes      XQL: No
\end{verbatim}
In LOREL and in XML-QL, both inspired to the SQL paradigm, queries can
be nested at an arbitrary level. In XML-GL and XQL nesting is not
supported. In XSL, templates can be nested.

\item {\em Set operations.} As in SQL, a query can be binary,
composed of the union, intersection, or difference of subqueries.
\begin{verbatim}
LOREL: Yes   XML-QL: Partially   XML-GL: Yes   XSL: Yes     XQL: Yes
\end{verbatim}
LOREL supports union, difference, and intersection.  XML-QL supports
union and intersection but has no difference.  In XML-GL, queries
allow multiple graphs in the left side of the query; this gives the
expressive power of union. Negation gives to XML-GL the expressive
power of a difference, and intersection can be built by repeatedly
applying negation.
XSL admits union and negation in the pattern language; there is not 
explicit intersection.
XQL supports union and intersection; negation gives to 
XQL the expressive power of a difference.

\end{enumerate}

\subsubsection*{G. Order Management}
\begin{enumerate}
\item{\em Ordering the result.} Consists of ordering the element
instances according to the ascending or descending values of some data
of the result, as performed by the {\tt order by} clause in SQL.
\begin{verbatim}
LOREL: Yes    XML-QL: Yes    XML-GL: Yes    XSL: Yes      XQL: No
\end{verbatim}
LOREL and XML-QL may order the result is gained through the {\tt
order by} clause.  In XML-GL, a leaf node can be labelled with {\tt
ASC}, {\tt DESC ORDER} and the elements extracted are in ascending or
descending order with respect to that node; multiple labelings are 
possible. In XSL,
sorting is specified by adding {\tt xsl:sort} elements as children of 
{\tt xsl:apply-templates} or {\tt xsl:for-each}. Each {\tt xsl:sort} 
specifies a sort key: the elements are ordered according to the specified 
sort keys, instead of being ordered on the basis of the document order 
(default).
In XQL, (as from available documentation) there are no clauses concerning 
the order of the result.
\item {\em Order-preserving result.} Consists of ordering the elements
in the result in the same way as the original document.
\begin{verbatim}
LOREL: Yes    XML-QL: Yes    XML-GL: Yes    XSL: Yes      XQL: Yes
\end{verbatim}     
In LOREL, the {\tt order by document order} means that the
retrieved elements are ordered in the same way as the original
document, and the new ones are placed at the end of the document with
an unspecified order among them; otherwise, the order of elements in
the result is left unspecified. XML-QL, XSL and XQL produce ordered XML,
therefore the order can be specified to be the same as the source
document (but such an order must be known).  XML-GL produces an
ordered XML element when one of the edges is marked (then, the
elements are produced in counterclockwise order w.r.t. the marked
edge).
\item {\em Querying the schema order.} Consists of asking for XML elements
and$\backslash$or attributes in a given order relationship, as they are 
specified in the schema. 
\begin{verbatim}
LOREL: No    XML-QL: Yes    XML-GL: No    XSL: No      XQL: No
\end{verbatim}
In XML-QL it is possible to query about the relative position of tags
within documents (asking if a tag precedes or follows another tag); this is 
not supported by the other languages. 

\item {\em Querying the instance order.} Consists of asking for XML 
elements and$\backslash$or attributes in a given order 
relationship, as they are appear in the instance of the document.
This is accomplished adding to the language the range qualifier operator, 
that enables the selection of a single number (or of a set of numbers) 
instances. 
\begin{verbatim}
LOREL: Yes    XML-QL: No    XML-GL: No    XSL: Yes      XQL: Yes
\end{verbatim}
In Lorel, the operator ``[$<$range$>$]'' is introduced into both the query 
and update language and it is applied against both a path expression or 
variable. 
In XML-QL and XML-GL, it seems that there is no range selection.
In XSL pattern language and in XQL, a specific node within a set of nodes is 
extracted simply enclosing the index ordinal within square brackets 
in the pattern. 
\end{enumerate}

\subsubsection*{H. Typing $\&$ Extensibility}
\begin{enumerate}
\item {\em Support of abstract data types.}  This feature concerns the
necessity of embedding inside an XML query language specialized
operations, i.e for selecting different kinds of multimedia content.
\begin{verbatim}
LOREL: Yes    XML-QL: No    XML-GL: No    XSL: No       XQL: No
\end{verbatim}
LOREL supports audio, video, images, and specialized data types such
as Jpeg, Gif, and Ps.

\item {\em Type coercion.} This feature concerns the ability of
implicit data casting among different types, as well as the ability to
compare values represented with different type constructors (e.g.,
scalars, singleton sets , and lists with one only element).
Because of the nature of semistructured data, the type coercion
provided by an XML query language should be much more flexible than
that of a database query language.  
\begin{verbatim}
LOREL: Yes    XML-QL: No    XML-GL: No    XSL: No       XQL: Partially
\end{verbatim}  

In LOREL, comparison between objects and$\backslash$or values is forced to
do ``the most intuitive thing'' when comparing objects and values of
different types. Coercion rules are provided for the various atomic types
and the corresponding predicates or functions. Furthermore, a
comparison between atomic objects, complex objects, and sets of objects
is accepted when there is an obvious interpretation.

In XQL, two values are comparable only after explicit casting of them,
as happens in a traditional programming language. The other languages
do not mention type coercion.

\end{enumerate}

\subsubsection*{I. Integration with XML}
\begin{enumerate}
\item {\em Support of RDF and/or XML Schemas.} RDF and XML Schemas are
emerging standards for representation of metadata regarding XML
documents and it may be desiderable to embed them into the query
language.
\begin{verbatim}
LOREL: No    XML-QL: No    XML-GL: No    XSL: No      XQL: No
\end{verbatim}

\item {\em Support of XPointer and XLink}. Intra-document and
inter-document linking could influence the evolution of the XML query
language; a query should be able to give as result not only XML data,
but also XML pointers and links.
\begin{verbatim}
LOREL: No    XML-QL: No    XML-GL: No    XSL: No      XQL: No
\end{verbatim}

\item {\em Support of tag variables.} This feature concerns the
possibility of explicitly querying the tag name rather than the tag
content.
\begin{verbatim}
LOREL: Yes    XML-QL: Yes    XML-GL: No    XSL: No       XQL: No
\end{verbatim}
In LOREL, path expressions return into variables of a special {\em path}
type the fully specified names of all paths that are reachable from
the path expressions themselves. Such names can be used
for building the names of the query result using the {\tt unquote()}
function. In XML-QL, variables can be
associated
to tags and then used to generate the tags of the result.
\end{enumerate}

\subsubsection*{J. Update language}
\begin{enumerate}
\item {\em Support for insert, delete, and update of elements.}
\begin{verbatim}
LOREL: Yes     XML-QL: No    XML-GL: Yes    XSL: No        XQL: No
\end{verbatim}
LOREL has an {\tt update} language. It is possible to create and
delete object names, create a new atomic or complex object, and modify
the value of an existing object.  Update operations in XML-GL are
graphically indicated as arrows labelled with {\tt I, D, U}. Each
primitive (except for delete that does not need the RHS) has a LHS
element and a RHS graph: the LHS is the target element of the
operation and the RHS graph represents the values to be inserted or
replaced by the primitive.  In XML-QL, XSL and XQL there is no update
language.
\end{enumerate}

\section{Comparative examples}
\label{examples}
In the following, we present a comparison of LOREL, XML-QL, XML-GL and
XQL on the basis of the query examples originally proposed by David
Maier in a Position Paper ``Database Desiderata for an XML Query
Language'' \cite{Mai98}; a preliminary version of the XML-QL examples were 
originally presented by Peter Fankhauser in a message to the XML 
Query language mailing list (message of Dec 22, 1998).  The underlying 
case study is that of a car dealer
office, with documents from different auto dealers and brokers.  The
{\tt manufacturer} documents list the manufacturer's name, year, and
models with their names, front rating, side rating, and rank; the
{\tt vehicle} documents list the vendor, make, year, color and price.
We consider XML data of the form:

{\small
\begin{verbatim}
<manufacturer>
<mn_name>Mercury</mn_name>
<year>1999</year>
<model>  <mo_name>Sable LT</mo_name>
<front_rating>3.84</front_rating>
<side_rating>2.14</side_rating>
<rank>9</rank>
</model>
....
</manufacturer>

<vehicle>
<vendor>Scott Thomason</vendor>
<make>Mercury</make>
<model>Sable LT</model>
<year>1999</year>
<color>metallic blue</color>
....
<price>26800</price>
</vehicle>
\end{verbatim}
}

\subsection{Query 1: Selection and Extraction}

{\em We want to select and extract $<manufacturer>$ elements 
where some $<model>$ has $<rank>$ less or equal to 10.} 

{\bf 1A. LOREL}
{\small
\begin{verbatim}
select M
from nhsc.manufacturer M
where M.model.rank <=10
\end{verbatim}
}
{\bf 1B. XML-QL}
{\small
\begin{verbatim}
WHERE <manufacturer>
       <model>
         <rank>$r</rank>
       </model>
      </manufacturer> ELEMENT_AS $m IN 
            www.nhsc\manufacturers.xml,
      $r<=10
CONSTRUCT $m
\end{verbatim}
}
{\bf 1C. XML-GL}
See Figure~\ref{query1}:
\begin{figure}[htbp]
\centerline{\psfig{figure=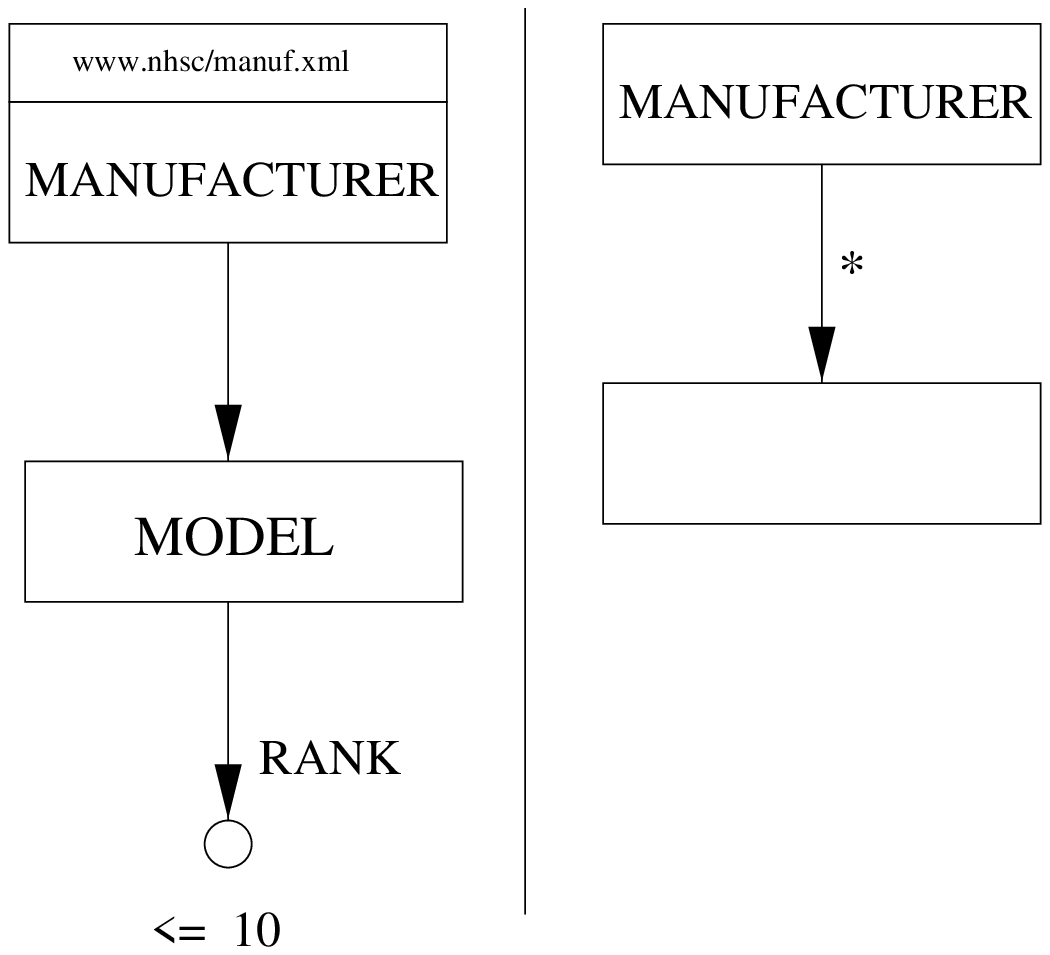,width=2in}}
\caption{Query 1}
\label{query1}
\end{figure}

{\bf 1D. XSL}
{\small
\begin{verbatim}
<xsl:template   match="/">
    <xsl:for each   select="manufacturer[model/rank<=10]">
          <xsl:value-of    />
    </xsl:for each>
</xsl:template>
\end{verbatim}    
}
{\bf 1E. XQL}
{\small
\begin{verbatim}
manufacturer[model/rank<=10]
\end{verbatim}
}
{\bf COMMENTS}
\begin{itemize}
\item In LOREL, the result is a collection of manufacturer object
identifiers. 

\item In XML-QL, the query applies to the XML
document {\tt www.nsch/manufacturers.xml}. It matches every
$<manufacturer>$ in the XML
document that has at least one
$<model>$, whose $<rank>$ is less or equal to 10. The presentation of
the result is a piece of XML document whose exact structure appears 
not to be defined in the references currently available.

\item In XML-GL, the query applies to the XML document 
{\tt www.nsch/manufacturers.xml}; it extracts all the occurrences of the
manufacturer elements satisfying the conditions stated in the 
LHS side. The elements used in the RHS to construct
the result are exactly those manufacturer objects retrieved in the LHS
with all the sub-elements as appearing in the input XML documents (but
without including the elements pointed by IDREFs links).  The result is a
new XML
document enclosed within the standard element {\tt result}.

\item In XSL, the rule applies to the root node and
the {\tt xsl:for-each} directive is 
instantiated for each {\tt manufac\-turer} node having at least one {\tt 
model} whose {\tt rank} is less or equal to 10. Through the {\tt 
xsl:value-of} instruction a text node is included in the result tree
for each selected {\tt manufacturer} element.

\item XQL does the job pretty concisely, having a navigation pattern
with a filter condition on the $<rank>$. The filter is
existentially quantified. The result is conventionally enclosed within
a standard element named {\tt xql:result}.

\item Summary: All languages cover the proposed example.
\end{itemize}

\subsection{Query 2: Reduction}

{\em From the $<manufacturer>$ elements, we want to drop those $<model>$
sub-elements whose $<rank>$ is greater than 10. We also want to elide the
$<front\_rating>$ and $<side\_rating>$ elements from the remaining
models.}

{\bf 2A. LOREL}
{\small
\begin{verbatim}
select Z.mn_name, Z.year,
     (select Z.model.mo_name, Z.model.rank
      where  Z.model.rank <= 10)
from nhsc.manufacturer Z
\end{verbatim}
}
{\bf 2B. XML-QL}
{\small
\begin{verbatim}
WHERE <manufacturer>
       <mn_name>$mn</mn_name>
       <year>$y</year>
      </manufacturer> CONTENT_AS $m IN 
            www.nhsc\manufacturers.xml
CONSTRUCT 
      <manufacturer>
         <mn_name>$mn</mn_name>
         <year>$y</year>
         {  WHERE <model>
                    <mo_name>$mon</mo_name>
                    <rank>$r</rank>
                  </model> IN $m,
                  $r<=10
            CONSTRUCT<model>
                       <mo_name>$mon</mo_name>
                       <rank>$r</rank>
                     </model>
         }
      </manufacturer>
\end{verbatim}
}
{\bf 2C. XML-GL}
See Figure~\ref{query2}:
\begin{figure}[htbp]
\centerline{\psfig{figure=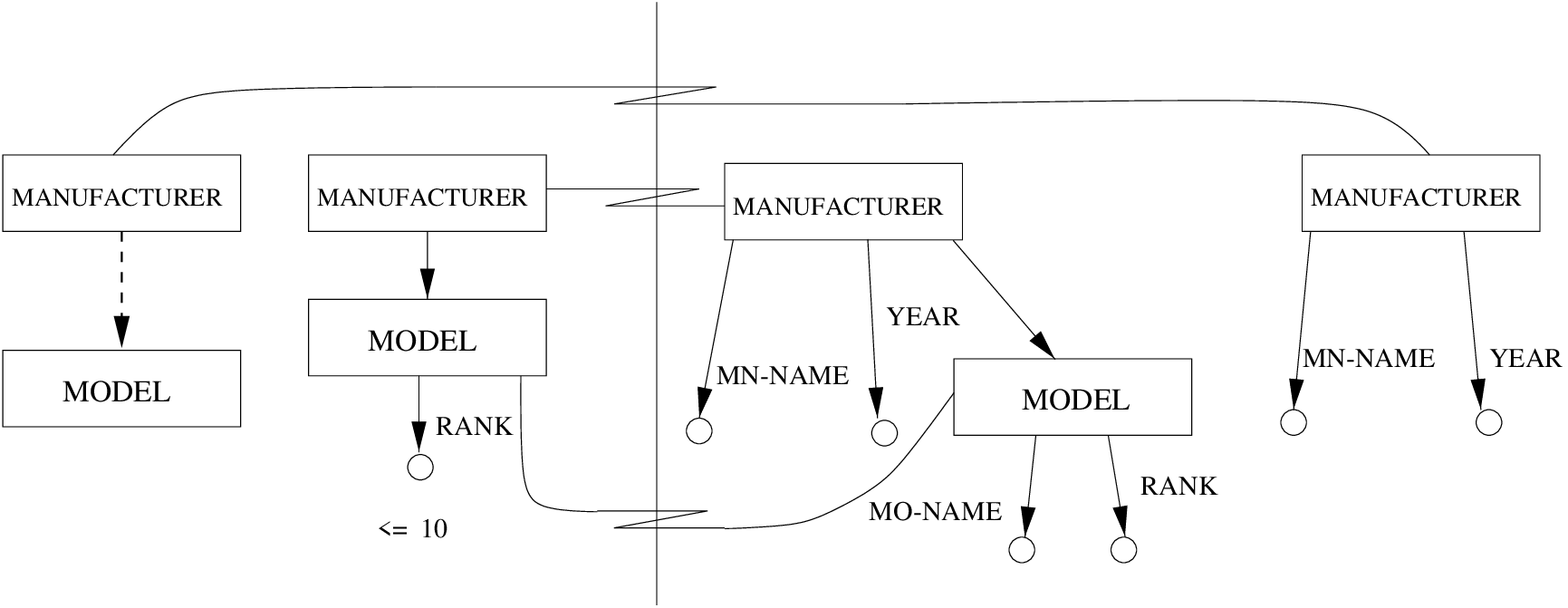,width=4in}}
\caption{Query 2}
\label{query2}
\end{figure}

{\bf 2D. XSL}
{\small
\begin{verbatim}
<xsl:template   match="manufacturer[model/rank<=10]">
   <model>
      <xsl:value-of   select="mo-name"/>
      <xsl:value-of   select="rank"/>
   </model>
</xsl:template>    
\end{verbatim}      
}
{\bf 2E. XQL}
{\small
\begin{verbatim}
Cannot be expressed.
\end{verbatim}
}
{\bf COMMENTS} 

\begin{itemize}
\item In LOREL the query consists of two nested subqueries one inside
the other; both are existentially quantified.

\item Also in XML-QL, the job is performed by nesting two subqueries;
nesting occurs within the construct clause of the first query.

\item XML-GL has no nesting, so the query 
selects first the elements of $<manufacturer>$ that do not have a
$<model>$ sub-element and puts them in the result; then, it selects those
elements having at least one $<model>$ element with suitable $<rank>$
value and inserts them in the result, by including only the selected
models.

\item In XSL, the template rule matches the {\tt manufacturer} elements 
satisfying the condition and, then, constructs the new {\tt model} elements 
with the subelements {\tt mo-name} and {\tt rank}.

\item In XQL, the query cannot be expressed neither as reduction nor as  
construction, because XQL does not allow restructuring, nested queries
or joins.

\item Summary: This example indicates that current query languages
lack of one relevant feature: reduction.  All languages must resort to a
solution based on the construction of the ``remainder'' of the
document, rather than eliding some elements.  This is possible only if
the DTD of the document is known.  XQL does not support construction,
so it cannot use this solution.
\end{itemize}

\subsection{Query 3: Joins}

{\em We want our query to generate pairs of $<manufacturer>$ and
$<vehicle>$ elements where $<mn\_name>$ = $<make>$, $<mo\_name>$ =
$<model>$ and $<year>$ = $<year>$.}

{\bf 3A. LOREL}
{\small
\begin{verbatim}
temp:= select (M,V) as pair
       from nhsc.manufacturer M, nhs.vehicle V
       where M.mn_name = V. make
       and   M.model.mo_name = V.model
       and   M.year = V.year
\end{verbatim}
}
{\bf 3B. XML-QL}
{\small
\begin{verbatim}
WHERE <manufacturer>
       <mn_name>$mn</mn_name>
       <year>$y</year>
       <model>
           <mo_name>$mon</mo_name>
       </model> CONTENT_AS $mo
      </manufacturer> CONTENT_AS $m IN 
            www.nhsc\manufacturers.xml
      <vehicle>
         <model>$mon</model>
         <year>$y</year>
         <make>$mn</make>
      </vehicle> CONTENT_AS $v IN www.nhsc\vehicles.xml     
CONSTRUCT
      <manufacturer>
         <mn_name>$mn</mn_name>
         <year>$y</year>
         <vehiclemodel>
            $mo,$v
         </vehiclemodel>
      </manufacturer>
\end{verbatim}
}
{\bf 3C. XML-GL}
See Figure~\ref{query3}:
\begin{figure}[htbp]
\centerline{\psfig{figure=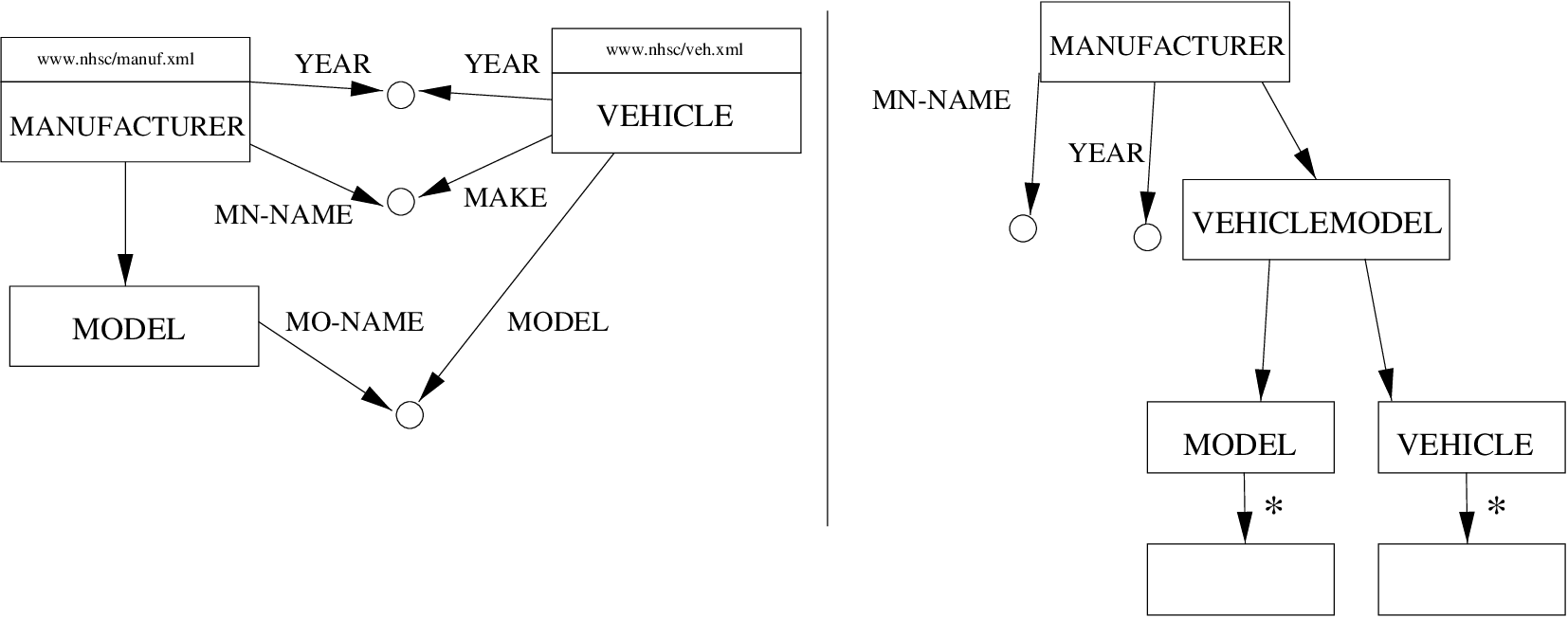,width=4in}}
\caption{Query 3}
\label{query3}
\end{figure}

{\bf 3D. XSL}
{\small
\begin{verbatim}
Cannot be expressed.
\end{verbatim}      
}
{\bf 3E. XQL}
{\small
\begin{verbatim}
Cannot be expressed.
\end{verbatim}
}
{\bf COMMENTS}

\begin{itemize}
\item In LOREL, the join builds pairs of OIDs of the relevant
documents after their joins. The joined elements are accessed creating a new 
entry point $temp$.

\item In XML-QL, a new piece of XML document is created and wrapped into the
tags $<vehiclemodel>$, with the content of the $<model>$ and
$<vehicle>$ elements that match on join conditions. 
  
\item In XML-GL, the pairs from $<model>$ and $<vehicle>$ are
extracted and made the
sub-elements of a new element named $<vehiclemodel>$, which is placed 
inside the $<manufacturer>$ element. 

\item In XSL, joins cannot be expressed.

\item XQL does not support joins.

\item Summary: Join is supported well by LOREL, XML-QL and XML-GL;
XSL and XQL lack the possibility of joining two documents.
\end{itemize}

\subsection{Query 4: Restructuring}
{\em We want our query to collect $<car>$ elements listing their make,
model, vendor, rank, and price, in this order.}

{\bf 4A. LOREL}
{\small
\begin{verbatim}
select xml(car: (select X.vehicle.make, X.vehicle.model,
                 X.vehicle.vendor, X.manufacturer.rank,
                 X.vehicle.price 
                 from temp.pair X))
\end{verbatim}
}
{\bf 4B. XML-QL}
{\small
\begin{verbatim}
WHERE <manufacturer>
       <mn_name>$mn</mn_name>
       <vehiclemodel>
          <model>          
          <mo_name>$mon</mo_name>
          <rank>$r</rank>
          </model> 
          <vehicle>
          <price>$y</price>
          <vendor>$mn</vendor>
          </vehicle> 
       </vehiclemodel>
      </manufacturer> IN www.nhsc\queryresult3.xml
CONSTRUCT
      <car>
         <make>$mn</make>
         <mo_name>$mon</mo_name>
         <vendor>$v</vendor>
         <rank>$r</rank>
         <price>$p</price>
      </car>

\end{verbatim}
}
{\bf 4C. XML-GL}
See Figure~\ref{query4}:
\begin{figure}[htbp]
\centerline{\psfig{figure=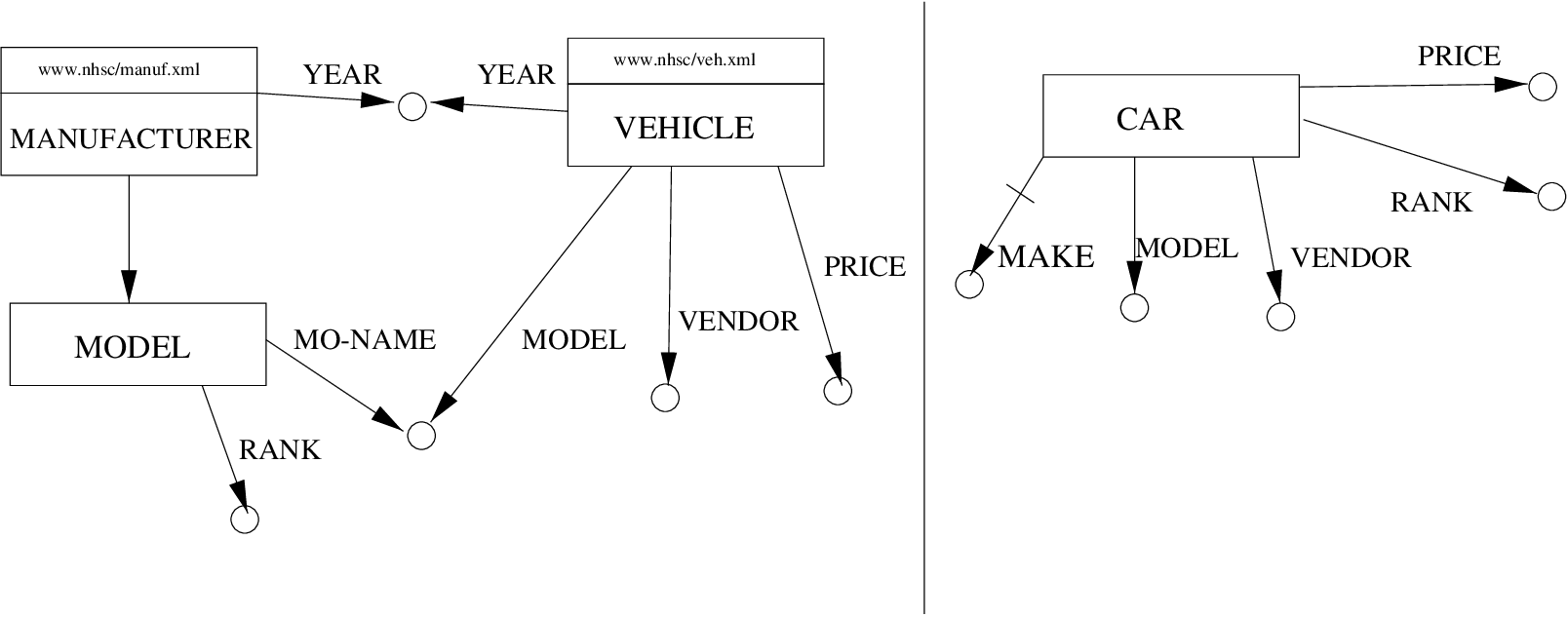,width=4in}}
\caption{Query 4}
\label{query4}
\end{figure}

{\bf 4D. XSL}
{\small
\begin{verbatim}
<xsl:template   match="manufacturer">
   <car>
      <xsl:value-of  select="vehiclemodel/vehicle/make"/>
      <xsl:value-of  select="vehiclemodel/model/mo-name"/>
      <xsl:value-of  select="vehiclemodel/vehicle/vendor"/>
      <xsl:value-of  select="vehiclemodel/model/rank"/>     
      <xsl:value-of  select="vehiclemodel/vehicle/price"/>    
   </car>
</xsl:template>
\end{verbatim}      
}
{\bf 4E. XQL}
{\small
\begin{verbatim}
Cannot be expressed.
\end{verbatim}
}
{\bf COMMENTS}

\begin{itemize}
\item In LOREL, the elements are extracted in the order in which they
appear in the query. The query result is associated with an element 
named $car$ by invoking the function $xml(car: query string)$.    
\item XML-QL deals with the ordering of the result explicitly
in the construct clause.
\item In XML-GL, a new $<car>$ element is introduced in the result and
enriched with the corresponding sub-elements, which are ordered
counterclock-wise by means of the graphic notation of marking one of
the edges.
\item In XSL, the template rule matches the {\tt manufacturer} elements and 
constructs in the result, the new {\tt car} tags filled with the appropriate 
sub-elements. 
\item In XQL, the query cannot be written, as it includes a 
join. 
\item Summary: this is a less interesting query, again it shows that join and
ordering 
are feasible with LOREL, XML-QL, XML-GL, while joins are unfeasible with
XSL and XQL.
\end{itemize}

\section{Desired Qualities}
\label{qualities}
Besides the comparative review of features, summarized in Table
\ref{table1}, we can briefly discuss the
desired qualities of query languages, and propose a relative ranking
of the five languages.

\begin{figure*}[h]
\centerline{\psfig{figure=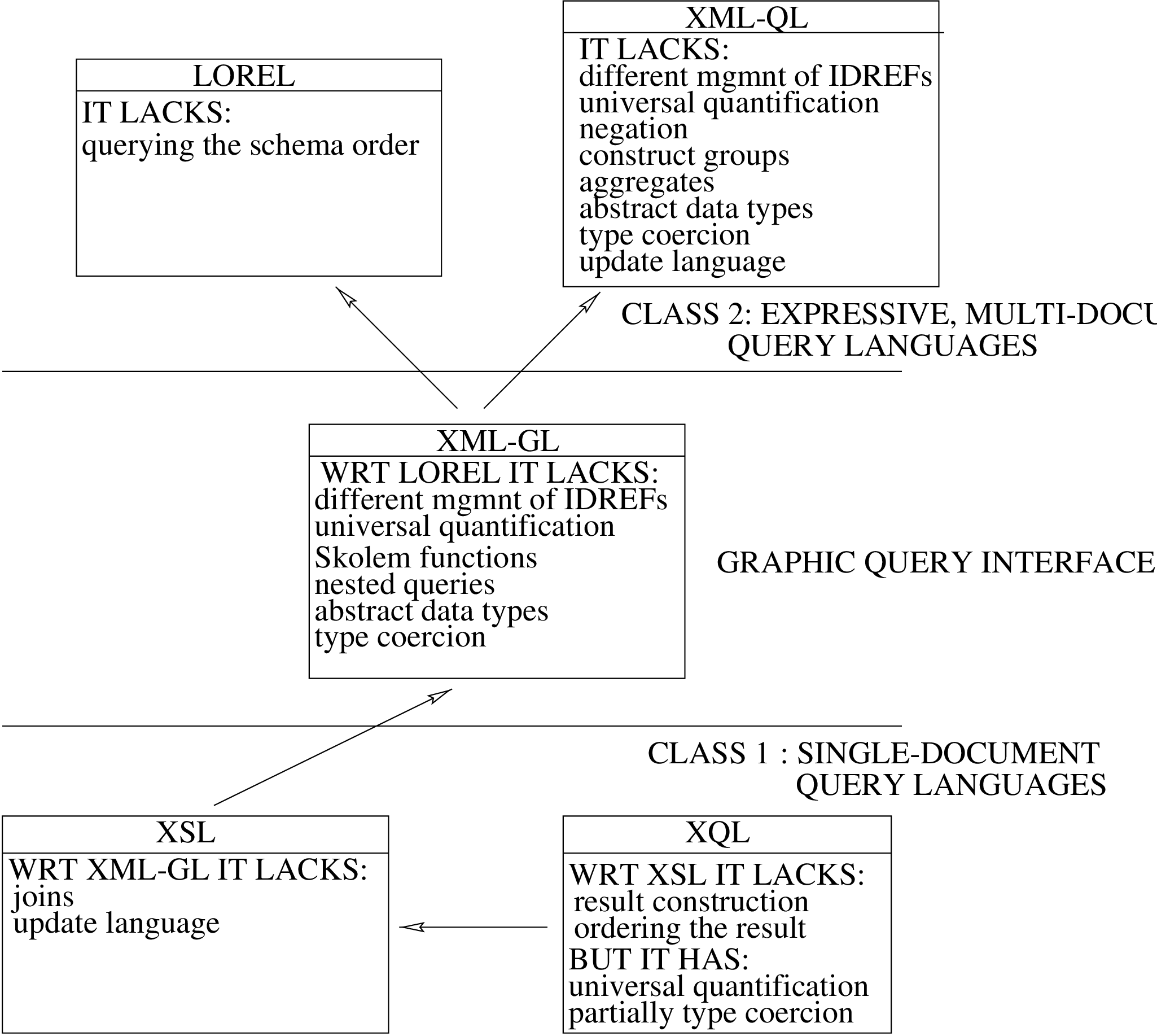,width=4in}}
\caption{Summarization of the expressive power of the languages.\newline
None of the five languages supports: Reduction, RDF embedding, 
XPointer and XLink embedding} 
\label{lack}
\end{figure*}

\begin{itemize}
\item{\sl Declarativeness.}
A query language is declarative when the specification of the query
defines the content of the result rather than a strategy to compute
it. According to the above definition, all the five query languages
are declarative. LOREL resembles OQL and therefore a calculus-based
language. XML-QL has variable unification as one of its ingredients
and as such recalls a logic language, although with a very peculiar
XML style. XML-GL recalls QBE. Both XSL and XQL make use of URL-like
patterns extensively.

\item{\sl Expressive power.} Indicates how powerful is the language in
expressing queries. From this perspective, XML-QL and LOREL are the
two most powerful languages. According to the features that we have
listed, LOREL is more powerful for what concerns the differential
management of IDREFs, universal quantification, negation, aggregate
management, abstract data types, type coercion, and updates; XML-QL
enables predicates on tag ordering
that are lacking in LOREL.  Both languages lack of a reduction
operation and of support for RDF, XPointers, and Xlinks, other
emerging W3C standards.  However, it appears that the missing features
could be easily added to both languages so as to make them equally
expressive. Then, the two languages would have comparable semantics
embedded within quite different syntaxes, as LOREL is OQL-like, while
XML-QL queries resamble XML documents; we may say that XML-QL is
XML-like.

XML-GL is less powerful than both the above languages, mainly because
of the limitations in supporting nested queries, certain specified
path expressions, and Skolem functions. However, the language supports
what is naturally specified in a graphical formalism, and can be
considered as a QBE-equivalent of the other two languages, in the
sense that also QBE is less expressive than SQL in the relational
world, however capturing the essential expressive power of relational 
languages.

XSL and XQL are quite comparable in their expressive power, and less 
expressive than the other languages.

XSL is a stylesheet language enabling a single document query language
with transformation capabilities. It lacks joins, update operations and 
Skolem functions. But it allows, with respect to XQL, restructuring of 
documents. 

Finally, XQL captures a restricted class of queries; its 
major limitations are the lack of support of joins and result
construction. The class of queries supported by XQL are those relative
to a single document being tested and projected out when the test is
satisfied.

A visualization of the expressive power of languages is given in 
Figure~\ref{lack}.

\item{\sl Ease of use.} This property indicates how easy it is to
write and/or to understand a query for an XML query programmer. From
this perspective, XML-GL is easy to read and to
understand, as it is associated to a graphic interface, while XQL is
also simple and easy, but much less powerful; these two
languages are the easiest to use. 
With respect to XQL, XSL is less readable because of the instruction tags 
and of the template rules. 
LOREL and XML-QL are comparable with
respect to the ease of formulation, although database experts will be
probably more familiar with the LOREL style, and XML experts will
probably be more familiar with XML-QL. One should also notice, e.g., from
the examples of Section 3, that XML-QL queries are rather verbose if
compared with LOREL queries.
\end{itemize}

\section{Conclusions: A Unified View}
\label{conclusions}
The five reviewed languages can be organized in a taxonomy, where:
\begin{itemize}
\item LOREL and XML-QL are the OQL-like and XML-like representatives
      of {\bf Class 2 of expressive query languages for XML}, playing 
      the same role as high-level SQL standards and languages (e.g., 
      SQL2) in the relational world. Our
      study indicates that they need certain additions in order
      to become equivalent in power, in which case it would be
      possible to translate between them. Currently, a major
      portion of the queries that they accept can be translated from 
      any language to another.
\item XSL and XQL are representative of {\bf Class 1 of single-document query
      languages}, playing the same role as core SQL standards and 
      languages (e.g., the SQL 
      supported by ODBC) in the relational world; as a major limitation, 
      we recall that they cannot join two different documents. 
      Their expressive power is included within the expressive
      power of Class 2 languages. Their rationale is to
      extract information from a single document, to be expressed 
      as a single string and passed as one of the
      URL parameters.
\item XML-GL can be considered as a {\bf graphical query interface to XML}, 
      playing the same role as graphical query interfaces (e.g., QBE) in the 
      relational world. The queries being 
      supported by XML-GL are the most relevant queries
      supported by Class 2 languages.
\end{itemize}
When the common features (as initially identified in this paper)
will become fully understood, it will be possible to envision a collection of
translators between languages of the same class, and/or between
languages of different classes, and/or from the graphic language
XML-GL to the programmative languages of Classes 1 and 2. In this
way, query languages for XML will constitute a language hierarchy
similar to the one existing for relational and object-relational
databases. 

\medskip
\begin{table*}[h]
\begin{center}
\begin{tabular} {|c|c|c|c|c|c|} \hline
 & LOREL & XML-QL & XML-GL & XSL & XQL  \\ \hline
{\small Specific data model}  & Yes & Yes & Yes & Yes & No   \\ \hline
{\small Differential mgmnt of IDREFs}  & Yes & No & No & No & No   \\ \hline
{\small Document selection}  & Yes & Yes & Yes & Yes & Yes   \\ \hline
{\small Joins} & Yes & Yes & Yes & No & No   \\ \hline
{\small Semantics of result} & {\small Set of OIDs} & {\small XML Doc}
& {\small XML Doc} & {\small XML Doc} & {\small XML Doc}\\ \hline
{\small Partially specified path expr.} & Yes & Yes & {\small Partially} &
Yes & Yes \\ \hline
{\small Halt on matching cyclic data} & Yes & {\small Undefined} & Yes &
No & {\small Undefined} \\ \hline
{\small Existential quantification} & Yes & Yes & Yes & Yes & Yes \\ \hline
{\small Universal quantification} & Yes & No & No & No & 
Yes   \\ 
\hline {\small Negation} & Yes & No & Yes & Yes & Yes   \\ \hline
{\small Reduction} & No & No & No & No & No \\ \hline
{\small Construct new elements} & Yes & Yes & Yes & Yes & No   \\ \hline
{\small Construct Groups} & Yes & No & Yes & No & No  \\ \hline
{\small Skolem functions} & Yes & Yes & {\small Partially} & No & No   \\ 
\hline
{\small Aggregates} & Yes & No & Yes & {\small Partially} & {\small 
Partially}  \\ 
\hline {\small Nested queries} & Yes & Yes & No & Yes & No    \\ \hline
{\small Set operations} & Yes & {\small Partially} & Yes & Yes & Yes  \\ 
\hline 
{\small Ordering the result} & Yes & Yes & Yes & Yes & No   \\ \hline
{\small Order-preserving result} & Yes & Yes & Yes & Yes & Yes   \\ \hline
{\small Querying the schema order} & No & Yes & No & No & No   \\ \hline
{\small Querying the instance order} & Yes & No & No & Yes & Yes \\ \hline 
{\small Abstract data types} & Yes & No & No & No & No   \\ \hline
{\small Type coercion} & Yes & No & No & No & {\small Partially}    \\ \hline
{\small Support of RDF} & No & No & No & No & No  \\ \hline
{\small Support of XPointer \& XLink} & No & No & No & No & No  \\ \hline
{\small Tag variables} & Yes & Yes & No & No & No\\ \hline
{\small Update language} & Yes & No & Yes & No & No  \\ \hline
\end{tabular}
\end{center}
\caption{Comparison table for XML query languages}
\label{table1}
\end{table*}

\section*{Acknowledgement}
We like to acknowledge the suggestions and criticisms of Serge
Abiteboul, Sara Comai, Jason McHugh, Yannis Papakonstantinou, Dan Suciu, 
Letizia Tanca and Jennifer Wi\-dom to an early presentation of this survey.
We are particularly grateful to Jennifer and Jason for their extensive 
advices and significant contributions.


\begin{thebibliography}{ABI99}

\bibitem[Abi99]{Abi99} S. Abiteboul. On Views and XML.
In {\em Proc. of ACM SIGMOD/SIGACT Conf. on Princ. of Database Syst. (PODS)},
Philadelphia, PA, May-June 1999.

\bibitem[AG*97]{AG*97}S. Abiteboul, R. Goldman, J. McHugh, V. Vassalos,
and Y. Zhuge. Views for Semistructured Data. In {\em Proc. of the 
Workshop on Management of Semistructured Data}, Tucson, Arizona, May 1997. 

\bibitem[AQ*97]{AQ*97}S. Abiteboul, D. Quass, J. McHugh, J. Widom, and J.
Wiener. The Lorel Query Language for Semistructured Data. In 
{\em International Journal on Digital Libraries}, 1(1):68-88, April 1997. 


\bibitem[C*99]{C*99}  S. Ceri, S. Comai, E. Damiani, P. Fraternali,
S. Paraboschi
and  L. Tanca. XML-GL: a Graphical Language for Querying and Restructuring
WWW Data. In {\em Proc. of 8th Int. World Wide Web Conference}, WWW8, 
Toronto, Canada, May 1999,
http://www8.org/fullpaper.html.

\bibitem[Cl99a]{Cl99a} J. Clark. XSL Transformations (XSLT specification), 
1999, http://www.w3.org/TR/WD-xslt.

\bibitem[Cl99b]{Cl99b} J. Clark. Xml Path Language (XPATH), 1999,
http://www.w3.org/TR/xpath. 

\bibitem[DeR98]{DeR98}S. J. DeRose. XQuery: A Unified Syntax for Linking
and Querying General XML. In {\em Proc. of the Query 
Languages Workshop
(QL98)}, Cambridge, Mass., Dec.98, http://www.w3.org/TandS/QL/
QL98/pp/xque\-ry.html.

\bibitem[DF*98]{DF*98} A. Deutsch, M. Fernandez, D. Florescu, Alon Levy
and D.Suciu. XML-QL: A Query Language for XML. In {\em Proc. of the Query
Languages
workshop(QL98)}, Cambridge, Mass., December 1998,
http://www.w3.org/TR/1998/NOTE-xml-ql-19980819/.

\bibitem[DF*99]{DF*99} A. Deutsch, M. Fernandez, D. Florescu, Alon Levy
and D.Suciu. A Query Language for XML.
In {\em Proc. of 8th Int. World Wide Web Conference}, WWW8,
Toronto, Canada, May 1999,
www8.org/fullpaper.html.

\bibitem[DOM98]{DOM98} Document Object Model (DOM) Level 1 Specification.
W3C recommendation, October 1998, http://w3.org/ TR/REC-DOM-Level-1/.

\bibitem[GMW99]{GMW99}R. Goldman, J. McHugh, and J. Widom. From
Semistructured Data to XML:Migrating the Lore Data Model and Query
Language. In {\em Proc.
of the 2nd International Workshop on the Web and Databases} (WebDB '99),
Philadelphia, Pennsylvania, June 1999. 

\bibitem[LP*99]{LP*99}B. Lud\"ascher, Y. Papakonstantinou, P. Velikhov and
V. Vianu. View Definition and DTD Inference for XML. In {\em Post-ICDT
Workshop on Query Processing for Semistructured Data and Non-Standard Data
Formats}, Jerusalem, 1999, 
http://www-rodin.inria.fr/external/ssd99/workshop.html.

\bibitem[Mai98]{Mai98}D. Maier. Database desiderata for an XML Query
Language. In
{\em Proc. of the Query Languages workshop}, Cambridge, Mass., Dec.
1998, http://www.w3.org/TandS/QL/QL98/pp/maier.h\-tml.

\bibitem[Qua98]{Qua98}D. Quass. Ten Features Necessary for an XML Query
Language. In
{\em Proc. of the Query Languages workshop}, Cambridge, Mass., Dec.
1998, http://www.w3.org/TandS/QL/QL98/pp/quass.h\-tml.

\bibitem[RLS98]{RLS98} J. Robie, J. Lapp and D. Schach. XML Query Language
(XQL). In
{\em Proc. of the Query Languages workshop}, Cambridge, Mass., Dec.
1998, http://www.w3.org/TandS/QL/QL98/pp/xql.html.

\bibitem[Ro99a]{Ro99a} J. Robie. XQL FAQ. http://metalab.unc.edu/xql/.

\bibitem[Ro99b]{Ro99b} J. Robie. XQL Tutorial. 
http://metalab.unc.edu /xql/xql-tutorial.html.

\bibitem[Ro99c]{Ro99c} J. Robie. The design of XQL.
http://www.w3.org /Style/XSL/Group/1998/09/XQL-design.html.

\bibitem[SLR98]{SLR98} D. Schach, J. Lapp, J. Robie. Querying and
Transforming
XML. In {\em Proc. of the Query Languages workshop}, Cambridge, Mass.,
Dec.1998, http://www.w3.org/TandS/QL/QL98/pp/query-transform.html.

\bibitem[Suc99]{Suc99} D. Suciu. Managing Web Data. A Tutorial presented
at {\em ACM SIGMOD/SIGACT Conference}, Philadelphia, PA, May-June 1999.

\bibitem[XML98]{XML98} XML 1.0. W3C
recommendation, February 1998, http://www.w3.org/TR/REC-xml.


\bibitem[XSL99]{XSL99} XSL Specification.
W3C Working Draft, 21 Apr 1999, http://www.w3.org/TR/WD-xsl/. 

\bibitem[W*98]{W*98} W3C XSL Working Group. 
The Query Language Position Paper of the XSL Working Group. 
In {\em Proc. of the Query Languages workshop}, Cambridge, 
Mass.,Dec.1998, 
\nolinebreak[4]http://www.w3.org/TandS/QL/ QL98/pp/xsl-wg-position.html.

\end{thebibliography}
\end{document}